\documentclass[%
 reprint,
showpacs,
 amsmath,amssymb,
 aps,
 pre,
 twocolumn
]{revtex4-1}
\usepackage{graphicx}
\usepackage{color}
\usepackage{dcolumn}
\usepackage{bm}
\usepackage{hyperref}

\newcommand{\muR}{{\mu_\mathrm{R}}}
\newcommand{\muL}{{\mu_\mathrm{L}}}
\newcommand{\ec}{{\epsilon_\mathrm{c}}}
\newcommand{\etac}{{\eta_{\mathrm{c}}}}
\newcommand{\TR}{{T_\mathrm{R}}}
\newcommand{\TL}{{T_\mathrm{L}}}

\newcommand{\betaR}{{\beta_\mathrm{R}}}
\newcommand{\betaL}{{\beta_\mathrm{L}}}
\newcommand{\fR}{{f_\mathrm{R}}}
\newcommand{\fL}{{f_\mathrm{L}}}
\newcommand{\JQR}{{J_Q^{\mathrm{R}}}}
\newcommand{\JQL}{{J_Q^{\mathrm{L}}}}

\newcommand{\ANR}{{A_N^{\mathrm{R}}}}
\newcommand{\ANL}{{A_N^{\mathrm{L}}}}

\newcommand{\AQL}{{A_Q^{\mathrm{L}}}}
\newcommand{\tauLR}{{\tau_{\mathrm{L}\to\mathrm{R}}}}
\newcommand{\tauRL}{{\tau_{\mathrm{R}\to\mathrm{L}}}}
\newcommand{\dUL}{{\text{d}U^{\text{L}}}}
\newcommand{\dNL}{{\text{d}N^{\text{L}}}}
\newcommand{\dQL}{{\text{d}Q^{\text{L}}}}

\newcommand{\dUR}{{\text{d}U^{\text{R}}}}
\newcommand{\dNR}{{\text{d}N^{\text{R}}}}
\newcommand{\dQR}{{\text{d}Q^{\text{R}}}}

\newcommand{\dUa}{{\text{d}U^{\alpha}}}
\newcommand{\dNa}{{\text{d}N^{\alpha}}}
\newcommand{\dQa}{{\text{d}Q^{\alpha}}}
\newcommand{\dWa}{{\text{d}W^{\alpha}}}

\newcommand{\SRdot}{{\dot{S}_\mathrm{R}}}
\newcommand{\SLdot}{{\dot{S}_\mathrm{L}}}
\newcommand{\QRdot}{{\dot{Q}_\mathrm{R}}}
\newcommand{\QLdot}{{\dot{Q}_\mathrm{L}}}
\newcommand{\QTdot}{{\dot{Q}_\mathrm{T}}}

\newcommand{\Sdot}{{\dot{S}}}
\newcommand{\TT}{{T_\mathrm{T}}}
\newcommand{\muT}{{\mu_\mathrm{T}}}
\newcommand{\JQT}{{J_Q^{\mathrm{T}}}}

\begin{document}

\title{Thermodynamics of the mesoscopic thermoelectric heat engine beyond the linear-response regime}

\author{Kaoru Yamamoto}
 \email{kaoru3@iis.u-tokyo.ac.jp
}
\affiliation{%
Department of Physics, The University of Tokyo, Komaba, Meguro, Tokyo 153-8505
, Japan}%

\author{Naomichi Hatano}
 \email{hatano@iis.u-tokyo.ac.jp
}

\affiliation{
Institute of Industrial Science, The University of Tokyo, Komaba, Meguro, Tokyo 153-8505, Japan}%

\date{\today}

\begin{abstract}
Mesoscopic thermoelectric heat engine is much anticipated as a device that allows us to utilize with high efficiency wasted heat inaccessible by conventional heat engines. 
However, the derivation of the heat current in this engine seems to be either not general or described too briefly, even inappropriate in some cases.
In this paper, we give a clear-cut derivation of the heat current of the engine with suitable assumptions beyond the linear-response regime. 
It resolves the confusion in the definition of the heat current in the linear-response regime. 
After verifying that we can construct the same formalism as that of the cyclic engine, we find the following two interesting results within the Landauer-B\"uttiker formalism: the efficiency of the mesoscopic thermoelectric engine reaches the Carnot efficiency if and \textit{only} if the transmission probability is finite at a specific energy and zero otherwise; the unitarity of the transmission probability guarantees the second law of thermodynamics, invalidating Benenti \textit{et al}.'s argument in the linear-response regime that one could obtain a finite power with the Carnot efficiency under a broken time-reversal symmetry. These results demonstrate how quantum mechanics constrains thermodynamics.
\end{abstract}


\maketitle

\section{Introduction}
Thermoelectric heat engine \cite{humphrey2002reversible,humphrey2005reversible,vendenbroeck2005thermodynamic,
esposito2009thermoelectric,esposito2009universality,natthapon2010thermoelectric,
gaveau2010stochastic,sanchez2011optimal,benenti2011thermodynamic, saito2011thermopower, sothmann2012rectification, kennes2013efficiency, balachandran2013efficiency, brandner2013strong,brandner2013multi, benenti2013fundamental,  hershfield2013nonlinear,whitney2013thermodynamic, whitney2013nonlinear,benenti2013conservation,mazza2014thermoelectric,
bergenfeldt2014hybrid,sothmann2014quantum,whitney2014most,
brandner2015bound,sothmann2015thermoelectric, crepieux2015mixed,whitney2015finding,minchev2015energy,
sanchez2015chiral} is much anticipated as a device that allows us to utilize wasted heat inaccessible by  conventional heat engines. This engine operates in a nonequilibrium steady state and converts heat to useful electrical power steadily, so that we do not need nonsteady processes used in cyclic engines, such as adiabatic compression, isothermal expansion, and so forth. Its efficiency, however, so far has been too low to use in terms of the figure of merit $Z(T)$, which is a serious problem in the field of thermoelectricity \cite{dresselhaus2007new,snyder2008complex,shakouri2011recent}. 

The \textit{mesoscopic} thermoelectric heat engine \cite{humphrey2002reversible,humphrey2005reversible,
esposito2009thermoelectric,esposito2009universality,natthapon2010thermoelectric,
sanchez2011optimal,
 saito2011thermopower,sothmann2012rectification, kennes2013efficiency, balachandran2013efficiency, brandner2013strong,brandner2013multi, benenti2013fundamental,  hershfield2013nonlinear,whitney2013thermodynamic, whitney2013nonlinear,mazza2014thermoelectric,
bergenfeldt2014hybrid, sothmann2014quantum,whitney2014most,
brandner2015bound,sothmann2015thermoelectric, crepieux2015mixed,whitney2015finding,minchev2015energy,
sanchez2015chiral} has emerged as a possible solution.
This engine is expected to have high efficiency thanks to the potential of nanoscale thermoelectricity \cite{dresselhaus2007new,snyder2008complex,shakouri2011recent}.
Moreover, this engine can be a powerful tool to investigate how quantum mechanics affects thermodynamics. 
For example, it has been argued that the heat current may be bounded because of the uncertainty principle \cite{pendry1983quantum,whitney2013thermodynamic, whitney2014most, whitney2015finding} and that the unitarity of the scattering matrix may give a new bound for the Onsager coefficients in the linear-response regime \cite{brandner2013strong,brandner2013multi,brandner2015bound}. 

In order to calculate the efficiency of a mesoscopic heat engine, we need to know the expression of the heat current going into or from a mesoscopic system, such as a quantum dot and a quantum wire.
Although many researchers have already used the definition of the heat current in the linear-response regime \cite{saito2011thermopower,balachandran2013efficiency,brandner2013strong,brandner2013multi, benenti2013fundamental,mazza2014thermoelectric,
sothmann2014quantum, brandner2015bound,sanchez2015chiral} as well as in nonlinear-response regimes \cite{humphrey2002reversible,humphrey2005reversible,esposito2009thermoelectric,esposito2009universality,natthapon2010thermoelectric,sanchez2011optimal,sothmann2012rectification,kennes2013efficiency, hershfield2013nonlinear,whitney2013thermodynamic,whitney2013nonlinear,
meair2013scattering, lopez2013nonlinear,hwang2013magnetic,whitney2014most,bergenfeldt2014hybrid,zotti2014heat,
sothmann2015thermoelectric, crepieux2015mixed,whitney2015finding,minchev2015energy,luengo2015heat,arrachea2009energy,moskalets2012scattering}, they did not give its derivation or explanation in most of the papers. 
There are several papers in which they explain the heat current.
In Ref.~\cite{sivan1986multichannel}, for example, Sivan and Imry gave a kind of derivation, which is the only one in the linear-response regime as far as we know.
However, it is not clear from the present authors' point of view; the derived expression is also inappropriate in nonlinear-response regimes as we will show below.
In nonlinear-response regimes, although several authors described the heat current \cite{arrachea2009energy,moskalets2012scattering, zotti2014heat,minchev2015energy, luengo2015heat,whitney2015finding}, their explanations seem too brief to understand for readers who are not familiar with the mesoscopic heat engine.
Moreover, since the presentation was done mostly in the Landauer-B\"uttiker formalism, the general framework seems to be hidden.
 
\begin{figure}
\begin{center}
\includegraphics[width=7cm]{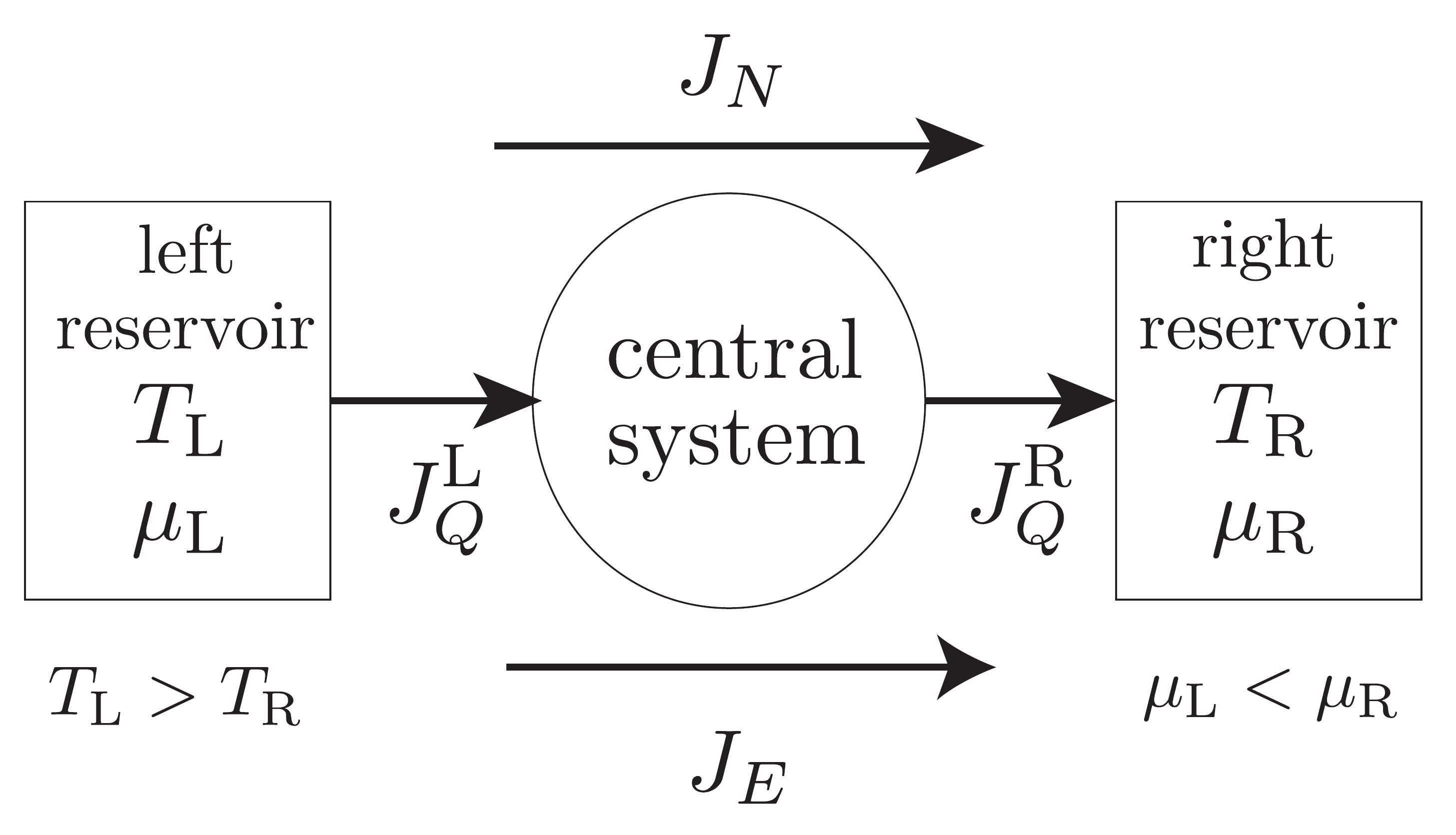}
\end{center}
\caption{Schematic picture of a mesoscopic heat engine.
We set the chemical potential of the right reservoir higher than the left, while the temperature of the left reservoir higher than the right so that an electric current may go from left to right against the difference of the chemical potential. 
}
\label{fig:engine}
\end{figure}

We, in the present paper, construct a thermodynamic formalism beyond the linear-response regime under reasonable assumptions. 
Since we do not use the Landauer-B\"uttiker formalism in the first half, our general formulation is applicable to systems with interactions as long as the assumptions are satisfied.
In Sec.~I\hspace{-.1em}I, we explain that thermoelectricity works as a heat engine under suitable conditions. 
In Sec.~I\hspace{-.1em}I\hspace{-.1em}I, after we briefly overview the confusion in the definition of the heat current in the linear-response regime, we first show the derivation of the heat currents  in the thermoelectric steady-state heat engine with suitable assumptions.
With the heat currents that we derived, we construct the general formalism.
In Sec.~I\hspace{-.1em}V, we give an example of the heat engine using the Landauer-B\"uttiker formalism, which we call the mesoscopic thermoelectric heat engine.
We can confirm the nonnegativity of the entropy production of this engine.
Moreover, we find that the efficiency of the mesoscopic thermoelectric engine reaches the Carnot efficiency if and \textit{only} if the transmission probability is finite at a certain energy and zero otherwise.
In Sec.~V, we give a model that expresses the situation when inelastic scatterings occur in the central system in Fig.~\ref{fig:engine}.
In Sec.~V\hspace{-.1em}I, we consider the system with a broken time-reversal symmetry using the Landauer-B\"uttiker formalism.
We find in nonlinear-response regimes that the unitarity of the transmission probability invalidates Benenti \textit{et al}.'s argument \cite{benenti2011thermodynamic} in the linear-response regime that one could obtain a finite power with the Carnot efficiency under a broken time-reversal symmetry.
It is remarkable that the unitarity, a quantum-mechanical concept, constrains thermodynamics; it guarantees that the Carnot efficiency is achieved only at zero power.

\section{Thermoelectric heat engine}
Let us explain the thermoelectric steady-state heat engine. 
Consider a central system, for example quantum dots or quantum wires, attached to two reservoirs on both sides; see Fig.~\ref{fig:engine}. 
We then make the following three assumptions: (i) the reservoirs are so much larger than the central system that it is always at equilibrium even if they interact with the central system, and hence we can define thermodynamic quantities of each reservoir, such as the temperature and the chemical potential; (ii) the central system has reached a nonequilibrium steady state in which there are constant flows; (iii) there is no entropy production in the central system because electrons undergo only elastic scatterings there. 

We can regard this system as a heat engine in the following situation.
We set the chemical potential of the right reservoir higher than that of the left, while the temperature of the left reservoir higher than that of the right.
We particularly set the reservoirs as well as the central system so that an electric current may go from left to right against the difference of the chemical potential. 
What happens per unit time is the following. 
Electrons gain heat $\JQL$ from the hot left reservoir, flow to the right against the potential difference, during which electrons do the work of amount 
\begin{equation}
\dot{W}=IV,
\end{equation}
where 
\begin{equation}
I=eJ_N
\end{equation}
is the electric current and 
\begin{equation}
V=\frac{\muR-\muL}{e}
\end{equation}
is the voltage difference, and then dump heat $\JQR$ to the cold right reservoir. 
We can thus consider this system as a heat engine.
This is an explanation specific to the case of electrons, which we can make more general as follows.
The central system gains heat from the left reservoir, does work, then dumps heat to the right reservoir. 
In this perspective, we can regard the central system as a working system of cyclic heat engines.
We will use this perspective hereafter. 
Because there is no entropy production in the central system, we must have 
\begin{equation}
\JQL -\JQR = \dot{W} . \label{eq:IV}
\end{equation}
Its efficiency $\eta$ is thereby given by
\begin{equation}
\eta = \frac{\dot{W}}{\JQL}=1-\frac{\JQR}{\JQL}, \label{eq:effc}
\end{equation}
which is the same as the standard cyclic heat engine.

Note that we can regard the system as a heat engine only when $J_N$ and $\JQL$ are positive.
The electrons do not necessarily flow from the hot reservoir to the cold one; the current which goes to the left is considered to be negative. 
The direction of the flow depends on $\muL$, $\muR$, $\TL$, and $\TR$ as well as details of the system. 
For example, we can regard the system as a refrigerator when electrons go from right to left against the temperature difference. In this case, the efficiency, which is called the coefficient of performance, is given by $\eta_{\text{cop}}=\JQL/(IV)$, not as in Eq.~\eqref{eq:effc}.

\section{Heat current of the thermoelectric heat engine}
\subsection{Confusion in the definition of the heat current in the linear-response regime}
Based on the above argument, we realize that we need \textit{two} heat currents $\JQL$ and $\JQR$ in order to discuss the thermoelectric heat engine properly.
Before deriving these heat currents, we mention the confusion in the definition of the heat current in the linear-response regime.

We first note that the energy current $J_E$ is often referred to as a ``heat'' current \cite{iyoda2010nonequilibrium,ruokola2011single}. This would be correct, though confusing, if the electrons did not do work and hence all energy became heat. This is certainly not correct in the situation in Fig.~\ref{fig:engine}.

Another definition $J_Q = J_E - \mu J_N$ was often used in the dawn of the research of the heat current in mesoscopic systems~\cite{sivan1986multichannel}. 
This definition may have been taken from Eq.~(17.8) in Callen's textbook~\cite{callen1985thermodynamics}.
Since this ``heat" current was not microscopically derived, we do not clearly know where it flows. It is indeed ambiguous of which part of the system in Fig.~\ref{fig:engine} the chemical potential $\mu$ of the heat current $J_Q = J_E - \mu J_N$ is. 
We should probably choose $\mu$ so that $J_Q$ may satisfy Onsager's reciprocal theorem. 
For example, in Ref.~\cite{sivan1986multichannel} the authors chose $\mu$ as $(\muL+\muR)/2$ and in Ref.~\cite{butcher1990thermal} the author chose $\mu$ as $\muL$. 
The choices do not make  any difference in the linear response of the voltage difference $V=(\muR-\muL)/e$ between $\JQL$ and $\JQR$ but differ in higher orders.

\subsection{Derivation of the heat currents going into or from a reservoir}
We here derive the heat currents $\JQL$ and $\JQR$ thermodynamically consistently. 
Although many researchers have used them, the derivation is often skipped or described too briefly.

Our starting point is to identify the heat currents coming into or going out of the central system as those coming from or going into the reservoirs.
We will give an important remark on this point later in this section.
We now derive the latter using general thermodynamics.
The first law of thermodynamics gives the relation
\begin{equation}
\dUa = \dQa + \dWa, \label{eq:firstlaw}
\end{equation}
where $\alpha=\text{L},\text{R}$, $\dUa$ and $\dQa$ denote the energy and heat flowing into the left or right reservoir, and $\dWa=\mu_\alpha\dNa$ is the work \textit{done to} the reservoir. 
Using Eq.~\eqref{eq:firstlaw}, we can express $\dQa$ in the form
\begin{equation}
\dQa = \dUa-\mu_\alpha\dNa. \label{eq:heat}
\end{equation} 
 
Since we treat the nonequilibrium steady state, we can define the changes of the particle number and the energy in a reservoir as a steady current, which enables us to define the particle and energy currents as 
\begin{equation}
J_N = -\frac{\dNL}{\text{d}t} = \frac{\dNR}{\text{d}t}, \  J_E = -\frac{\dUL}{\text{d}t } = \frac{\dUR}{\text{d}t}, \label{eq:JNJE}
\end{equation}
where we used the conservation of particle number and energy, $\dNL+ \dNR=0$ and $\dUL + \dUR=0$, because there is no dissipation in the central system. 
The negative signs appeared because we defined the positive direction so that currents going to the right may be positive. 
We then define the heat currents using Eqs.~\eqref{eq:heat} and \eqref{eq:JNJE} in the form
\begin{align}
\JQL &= -\frac{\dQL}{\text{d}t} = J_E - \muL J_N, \notag \\
\JQR &= \frac{\dQR}{\text{d}t} = J_E - \muR J_N \label{eq:JQ},
\end{align}
where $\JQL$ is the heat current flowing from the left reservoir into the system and $\JQR$ is that flowing from the system into the right reservoir. 

Note that these expressions of the heat currents are valid in the presence of many-body interactions if our assumptions (i) to (iii) hold. 
We can therefore apply them to systems for which we cannot use the Landauer-B\"uttiker formula.
We will also discuss a possible formulation in Sec.~V when there is an entropy production in the central system.
 
Let us come back to our starting point above.
It is crucial to note that the working system of the engine is now the central system in Fig.~\ref{fig:engine}, for which we cannot define the temperature and the chemical potential because it is highly nonequilibrium. 
In the textbook \cite{datta1997electronic}, Datta argued that we could define an ``effective" chemical potential in quantum wires, but it is in fact not a thermodynamic observable.
We should therefore be aware that it is not trivial at all to define the heat and the work for the central system.
We here identify the heat current coming into the central system as that coming  from a reservoir under the assumption below, which corresponds to the one that there is no entropy production in the central system.

According to the first law of thermodynamics, in order to calculate the heat current coming into or from the central system, we need to specify the energy current coming into or from it and the work done to or by it. 
The former is easy to specify as $J_E$ because the energy current is a conserved quantity. 
The problem arises when we try to specify the work done to or by the central system, in which we cannot define thermodynamic intensive variables.

We here make the following assumption to specify it: the work done to or by the central system is equal to that done by or to the reservoir. 
This assumption lets us find that the work done to the system on the left side is equal to that done by the left reservoir, $\muL J_N$, and the work done by the system on the right side is equal to that done to the right reservoir, $\muR J_N$.
We can thus find that the expression of the heat current coming into or from the central system is equal to those that we derived in Eq.~\eqref{eq:JQ}.
Although the expression of the heat currents corresponds to that used in previous researches, we believe that our derivation is more accurate than the previous ones.
The crucial point is that we derived the heat current coming into or from the central system, not a reservoir; we need to make the assumption above to specify it.
The assumption is probably equivalent to the assumption (iii) in Sec.~I\hspace{-.1em}I that there is no entropy production in the central system.

\subsection{General formalism of the thermoelectric heat engine}
Let us show that we can construct the same formalism as that of the cyclic engine using the heat currents  Eq.~\eqref{eq:JQ}.  We first easily find that they indeed satisfy Eq.~\eqref{eq:IV}.
This is the first check of the consistency of the expression Eq.~\eqref{eq:JQ}.
  
We then show that the upper limit of the efficiency Eq.~\eqref{eq:effc} is the Carnot efficiency as is expected from the theory of the standard cyclic heat engine.
Let   
\begin{equation}
\SLdot\text{d}t = \frac{\dQL}{\TL}
\end{equation}
and 
\begin{equation}
\SRdot\text{d}t = \frac{\dQR}{\TR}
\end{equation}
denote the entropy productions in the left and right reservoirs, respectively.
Using Eq.~\eqref{eq:JQ}, we can relate these entropy productions to the heat currents as 
\begin{equation}
\JQL =-\TL\SLdot, \quad \JQR = \TR\SRdot. \label{eq:JQLSdot}
\end{equation}
According to the second law of thermodynamics, an entropy production of an isolated system increases. Our whole system, which consists of the two reservoirs and the central system is indeed isolated. The net entropy production of the whole system 
\begin{equation}
\Sdot = \SLdot + \SRdot \label{eq:Sdot}
\end{equation}
is thus non-negative, that is, $-\SRdot \leq \SLdot$; we assumed that there is no entropy production in the central system.
We thus have from Eqs.~\eqref{eq:IV}, \eqref{eq:effc}, and \eqref{eq:JQLSdot},
\begin{equation}
\eta = \frac{-\TL\SLdot -\TR\SRdot}{-\TL\SLdot} \leq  1-\frac{\TR}{\TL}=\eta_{\text{c}}, \label{eq:effc3}
\end{equation}
where $\eta_{\text{c}}$ is the Carnot efficiency. 
We can achieve the equality if and only if $\SLdot =-\SRdot$, that is, $\Sdot = 0$.
We conclude that with the heat currents Eq.~\eqref{eq:JQ}, we can produce Eqs.~\eqref{eq:IV} and \eqref{eq:effc3}, which are the same formulas  as those of the cyclic engine.

\section{Heat current in the mesoscopic heat engine using the Landauer-B\"uttiker formula}
In order to justify the definitions Eq.~\eqref{eq:JQ} further, we here derive microscopic expressions for the \textit{mesoscopic} thermoelectric heat engine.
We additionally assume here that the central system in  Fig.~\ref{fig:engine} is a noninteracting coherent conductor that accommodates the Landauer-B\"uttiker formalism \cite{datta1997electronic}.
Note again that the arguments in Secs.~\\ I\hspace{-.1em}I and I\hspace{-.1em}I\hspace{-.1em}I are valid even in the presence of many-body interactions, to which the Landauer-B\"uttiker formalism is not applicable.

We can obtain $J_N$ and $J_E$  in Eq.~\eqref{eq:JQ} microscopically as
\begin{align}
 J_N &= \frac{1}{h}\int_{-\infty}^{\infty} d\epsilon \  \tau(\epsilon)(\fL(\epsilon)-\fR(\epsilon)), \label{eq:UL} \\
 J_E &= \frac{1}{h}\int_{-\infty}^{\infty} d\epsilon \  \tau(\epsilon)\epsilon(\fL(\epsilon)-\fR(\epsilon)),\label{eq:NL}
\end{align}
where $h$ is the Planck constant, 
\begin{equation}
f_{\alpha}(\epsilon) = \frac{1}{1+\exp(\beta_{\alpha}(\epsilon-\mu_{\alpha}))}
\end{equation}
is the Fermi distribution function of the reservoir ($\alpha=\text{L,R}$), $\beta_{\alpha}=T_{\alpha}^{-1}$ is the inverse temperature, and $\tau(\epsilon)$ is the transmission probability at energy $\epsilon$.
Substituting Eqs.~\eqref{eq:UL} and \eqref{eq:NL} into Eq.~\eqref{eq:JQ}, we arrive at 
\begin{equation}
J_Q^\alpha= \frac{1}{h}\int_{-\infty}^{\infty} d\epsilon  \ \tau(\epsilon)(\epsilon-\mu_\alpha)(\fL(\epsilon)-\fR(\epsilon)), \label{eq:jql}
\end{equation}
where $\alpha = \text{L},\text{R}$.

Note that these heat currents satisfy Onsager's reciprocal theorem when we expand them in terms of appropriate affinities; we can verify the theorem by expanding $J_N$ and $\JQL$ in terms of $\ANL = \betaL(\muL-\muR)$ and $\AQL = -(\betaL-\betaR)$ or $J_N$ and $\JQR$ in terms of $\ANR = \betaR(\muL-\muR)$ and $\AQL = -(\betaL-\betaR)$ \cite{yamamoto2015thermodynamics}.

We can also verify the non-negativity of the entropy production with the heat currents Eq.~\eqref{eq:JQ} and their microscopic expressions Eq.~\eqref{eq:jql}, although it must be satisfied anyway according to the second law of thermodynamics.
Using Eqs.~\eqref{eq:JQLSdot} and \eqref{eq:Sdot}, the total entropy production of the system is  generally expressed with the heat currents:
\begin{equation}
\Sdot = \SLdot + \SRdot = -\frac{\JQL}{\TL} + \frac{\JQR}{\TR}.
\end{equation} 
Substituting the microscopic expression of the heat currents Eq.~\eqref{eq:jql}, we obtain the microscopic expression of the total entropy production in the form
\begin{equation}
\Sdot = \int_{-\infty}^{\infty} d\epsilon \ \tau(\epsilon)(\fL(\epsilon) - \fR(\epsilon))\log\left[\frac{\fL(\epsilon)(1-\fR(\epsilon))}{\fR(\epsilon)(1-\fL(\epsilon))}\right] \label{eq:sdotintegrand} ,
\end{equation}
which Whitney derived \cite{whitney2013thermodynamic} starting from the nonlinear Landauer-B\"uttiker formalism.
Since $\tau(\epsilon) \ge 0$, the integrand in Eq.~(\ref{eq:sdotintegrand}) is always nonnegative, which leads to the nonnegativity of the entropy production.
The fact that we reproduced Whitney's microscopic expression Eq.~\eqref{eq:sdotintegrand} endorses our general expressions of the heat currents Eq.~\eqref{eq:JQ}.

Since we know that we can achieve the Carnot efficiency when the total entropy production of the whole system is zero, let us find the transmission probability that satisfies $\Sdot = 0$. 
We easily see the following condition: for each value of $\epsilon$, $\tau(\epsilon)=0$ or $\fL(\epsilon) - \fR(\epsilon) = 0$. If $\tau(\epsilon)=0$ for any $\epsilon$ or if $\fL(\epsilon) - \fR(\epsilon) = 0$ for any $\epsilon$, however, the transport would not happen, the engine would not work, and the efficiency $\eta$ would be trivially $0/0$. 
The only nontrivial condition is given by $\tau(\epsilon) \neq 0$ and  $\fL(\epsilon) - \fR(\epsilon) = 0$ at one value of energy; if we demanded $\fL(\epsilon) = \fR(\epsilon)$ at two values of energy, they would be equal at any values of energy.  

In order to achieve the above condition, the transmission probability should be of the form 
\begin{equation}
\tau(\epsilon) = \begin{cases}
    c & \text{for} \ \epsilon=\epsilon_c \\
   0 & \text{otherwise},
  \end{cases}  \label{eq:tau}
\end{equation}
where c is a constant satisfying $0<c\le1$ and
\begin{equation}
\epsilon_c = \frac{\TL\muR-\TR\muL}{\TL-\TR}
\end{equation}
is derived from the condition $\fL(\epsilon_c) = \fR(\epsilon_c)$.
Using Eqs.~\eqref{eq:effc}, \eqref{eq:JQ}, \eqref{eq:UL}, and \eqref{eq:jql}, we indeed find that this transmission probability is a necessary and sufficient condition for the Carnot efficiency:
\begin{equation}
\eta =  \frac{\muL-\muR}{\epsilon_c-\muL} = 1-\frac{\TR}{\TL}=\eta_{\text{c}}. \label{eq:etac}
\end{equation}

The transmission probability Eq.~\eqref{eq:tau} for the Carnot efficiency that we derived  is well known in the literature on  thermoelectricity \cite{mahan1996best,humphrey2002reversible,humphrey2005reversible,benenti2013fundamental,
hershfield2013nonlinear,whitney2014most, whitney2015finding}. 
However, in the mesoscopic thermoelectric engine with the Landauer-B\"uttiker formalism that we treat in the present paper, there has been no discussion that the transmission probability Eq.~\eqref{eq:tau} is a \textit{sufficient} condition for the Carnot efficiency \cite{benenti2013fundamental,
hershfield2013nonlinear,whitney2014most, whitney2015finding}, as far as we know.  
In contrast, we showed that the transmission probability Eq.~\eqref{eq:tau} is a necessary and \textit{sufficient} condition for the Carnot efficiency straightforwardly from the expression of the entropy production Eq.~\eqref{eq:sdotintegrand} along with the condition $\Sdot =0$.

\begin{figure}
\begin{center}
\includegraphics[width=5.5cm]{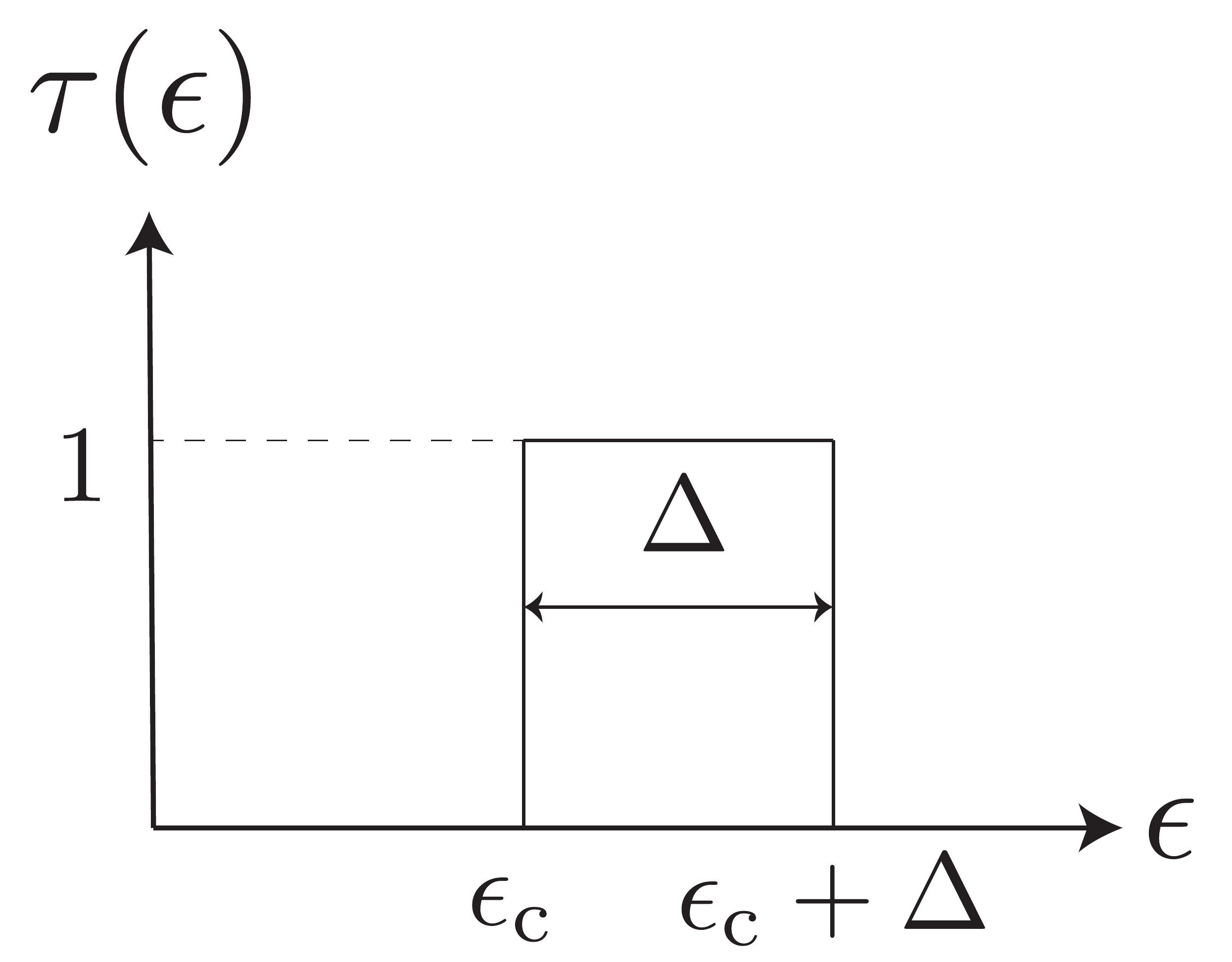}
\end{center}
\caption{The energy filter transmission probability.}
\label{fig:ewindow}
\end{figure}

In order to see what limit of this engine corresponds to the quasistatic limit of the cyclic heat engine, let us consider the transmission probability Eq.~\eqref{eq:tau} in the limit of $\Delta \to 0$ of the energy window \cite{whitney2014most, whitney2015finding} shown in Fig.~{\ref{fig:ewindow}}; 
we here set $c$ in Eq.~\eqref{eq:tau} as unity for simplicity.
The power is given by \cite{whitney2015finding}
\begin{align}
J_N(\muR-\muL) &= \frac{\muR-\muL}{h}\int_{\ec}^{\ec+\Delta}d\epsilon(\fL(\epsilon)-\fR(\epsilon)) \notag \\
 &= \frac{\muR-\muL}{2h}F(\ec)\Delta^2 + \mathcal{O}(\Delta^3),
\end{align}
where $F(\epsilon) = \partial/\partial \epsilon(\fL(\epsilon)-\fR(\epsilon))$. The heat current $\JQL = (\ec-\muL)F(\ec)\Delta^2/2h + \mathcal{O}(\Delta^3)$ vanishes and the efficiency $\eta = \eta_\mathrm{c} - \mathcal{O}(\Delta)$ becomes the Carnot efficiency in the limit of $\Delta \to 0$. 

The fact that the power vanishes in this limit is the same as in the standard heat engine; the Carnot cycle produces zero power because the quasistatic limit of the Carnot engine takes infinite time for us to operate the engine.
As $\Delta \to 0$, $J_N$ goes to zero, which means that the event that one particle transmits becomes rarer and rarer because $J_N$ is the mean value of the particle current. 
In other words, as $\Delta \to 0$, the mean transmission time of particles, that is, the average time that it takes for one particle to transmit, becomes longer and longer. 
We expect that there is a relation as $\Delta \sim \hbar t^{-1}$, where $t$ is the mean transmission time of particles. 
It thus takes infinite average time for particles to transmit from the left reservoir to the right reservoir in the limit $\Delta \to 0$. 
This physically corresponds to the quasistatic limit of the cyclic heat engine, in which it takes infinite time for us to operate the engine.

\section{The case of inelastic scattering}

\begin{figure}
\begin{center}
\includegraphics[width=7cm]{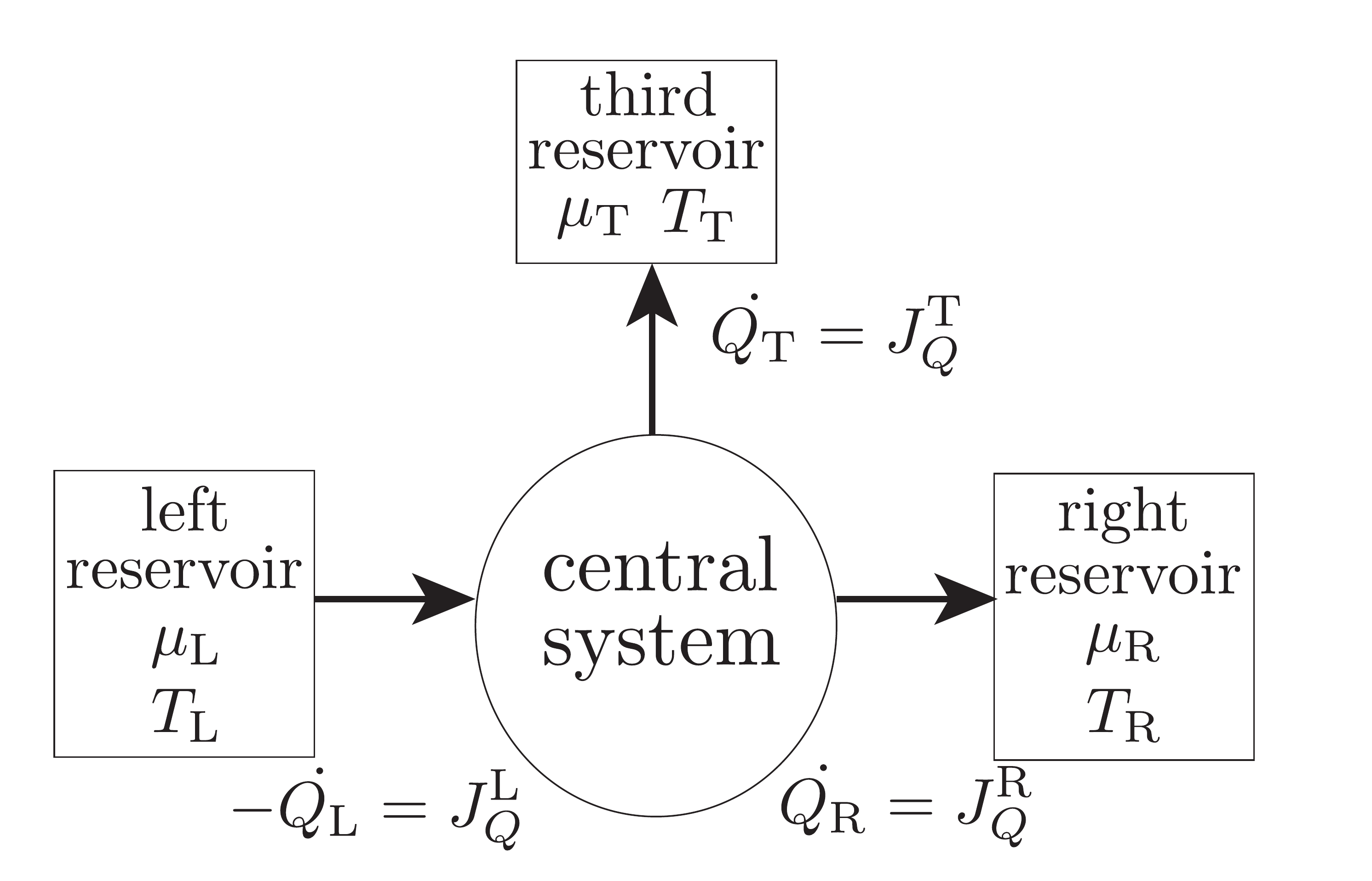}
\end{center}
\caption{A three-terminal model. The third bosonic reservoir on the top represents the energy dissipation.}
\label{fig:three}
\end{figure}

We here extend our theory to the case of inelastic scatterings, such as electron-hole and electron-phonon interactions, in which the number of electrons is conserved, but their energy is dissipated from the central system. In reality, the dissipated energy goes to an environment. 
In order to model this situation theoretically, we introduce a third bosonic reservoir (Fig.~\ref{fig:three}) \cite{entin2014enhanced}, into which the dissipated energy goes. 
By tuning its temperature $\TT$ and chemical potential $\muT$, we can make the net particle current going into the reservoir zero and the net energy current going into the reservoir positive. 
Note that the energy current going into the third reservoir is equal to the heat current going into it because the net particle current going into it is zero; see the expression of the heat current in Eq.~\eqref{eq:JQ}.
Let us denote $\QTdot$ as the heat current going into it hereafter.
By additionally tuning transmission probabilities going into the third reservoir from the other reservoirs, we should be able to set $\QTdot$ at any values observed in experiments.
Note, however, that this model cannot represent all cases of inelastic scatterings; for example, we may have to consider a non-Markov reservoir if the dissipation is non-Markov.
We here focus on the dissipation which we can represent with the model in Fig.~\ref{fig:three}.
We again let $\QLdot$ and $\QRdot$ denote the heat flowing per unit time into the left and right reservoirs, respectively, and $\dot{W}$ the work done by the central system per unit time. 
Note that we define the heat currents as $\QLdot = -\JQL$ and  $\QRdot=\JQR$ in the present paper.

The first law of thermodynamics then gives the following relation:
\begin{equation}
\QLdot+\QRdot+\QTdot + \dot{W} = 0.
\end{equation}
The efficiency is given by
\begin{equation}
\eta = \frac{\dot{W}}{-\QLdot} = \frac{\QLdot+\QRdot+\QTdot}{\QLdot}=\left(1+\frac{\QRdot}{\QLdot}\right) + \frac{\QTdot}{\QLdot}. \label{eq:3rdeff}
\end{equation}
By defining the dissipative heat current as $\QTdot = \JQT$, we can rewrite the efficiency Eq.~\eqref{eq:3rdeff} as 
\begin{equation}
\eta = \left(1-\frac{\JQR}{\JQL}\right) -\frac{\JQT}{\JQL} \le 1-\frac{\JQR}{\JQL}
\end{equation}
and obtain the work in the form
\begin{equation}
\dot{W} = (-\QLdot-\QRdot)-\QTdot = (\JQL-\JQR)-\JQT \le \JQL-\JQR .
\end{equation}
In the case of no dissipation $\JQT = 0$, the expressions of the efficiency and the work reduce to Eqs.~\eqref{eq:IV} and \eqref{eq:effc} in the case of two reservoirs, respectively.  
Since we consider the case in which $\QTdot = \JQT \ge0$, we find that the effect of the dissipation decreases the efficiency and the power. 
The entropy production of the \textit{whole} system is given as follows:
\begin{align}
\dot{S} &= \frac{\QLdot}{\TL}+\frac{\QRdot}{\TR}+\frac{\QTdot}{\TT} \notag \\
 &= -\frac{\JQL}{\TL}+\frac{\JQR}{\TR}+\frac{\JQT}{\TT}.
\end{align}
Because $\JQT\ge0$, the entropy production is still nonnegative if we introduce the third reservoir. 

\section{Second law, reciprocity, and unitarity}
We here remark on the relation of the entropy production with the reciprocity and the unitarity.
Using Onsager's reciprocal theorem, Benenti \textit{et al}.~\cite{benenti2011thermodynamic} recently proposed an interesting argument in the linear-response regime that a thermoelectric engine could have a finite power with the Carnot efficiency under a magnetic field. 
As this proposal is only in the linear-response regime, we hereafter argue its possibility in nonlinear-response regimes. 

For this purpose, let us first impose Onsager's reciprocal relation without the unitarity of the transmission probability.
Without the unitarity, however, we can choose any expressions of $J_N$ because particles are not conserved without the unitarity.
For example, suppose that we choose seemingly plausible expression of $J_N$ as
\begin{equation}
J_N = \int_{-\infty}^{\infty} d\epsilon( \tauLR(\epsilon,B)\fL(\epsilon) - \tauRL(\epsilon,B)\fR(\epsilon)). \label{eq:jntau}
\end{equation}
Here $\tauLR(\epsilon,B)$ is the transmission probability for the electrons from left to right at energy $\epsilon$ and $\tauRL(\epsilon,B)$ that for the electrons from right to left at energy $\epsilon$.
This expression of $J_N$ leads to the expression of $\JQL$ in the form
\begin{equation}
\JQL = \int_{-\infty}^{\infty} d\epsilon (\epsilon-\muL)(\tauLR(\epsilon,B)\fL(\epsilon) - \tauRL(\epsilon,B)\fR(\epsilon)). \label{eq:jqltau}
\end{equation}
Let us further choose the transmission probabilities as $\tauLR(\epsilon,B) = a(B)$ at $\epsilon = \epsilon_c$ and otherwise zero, as well as $\tauRL(\epsilon,B) = b(B)$ at $\epsilon = \epsilon_c$ and otherwise zero, where $a(B)$ and $b(B)$ are constants with $a(B) > b(B)$. 
Note that these choices are not prohibited under Onsager's reciprocal relation $\tauLR(\epsilon,B) = \tauRL(\epsilon,-B)$ \cite{datta1997electronic}.
With Eqs.~\eqref{eq:jntau} and \eqref{eq:jqltau}, we find that the efficiency
\begin{equation}
\eta = \frac{IV}{\JQL} = \frac{J_N(\muR-\muL)}{\JQL}
\end{equation}
becomes the Carnot efficiency $\eta = \etac$, whereas the power has a finite value $J_N(\muR-\muL) = (a(B)-b(B))(\muR-\muL)>0$. 
Therefore, we could obtain a finite power with the Carnot efficiency only under Onsager's reciprocal theorem, which is the nonlinear version of Benenti \textit{et al}.'s argument in the linear-response regime \cite{benenti2011thermodynamic}.

However, the unitarity of the transmission probability prohibits this situation.
The unitarity condition of the transmission probability yields $\tauLR(\epsilon,B)=\tauRL(\epsilon,B)$, which makes $a(B) = b(B)$ at any $B$, and hence we find no power at the Carnot efficiency.
This corresponds to the fact that Brandner \textit{et al}.~\cite{brandner2013strong} found a new bound among Onsager's coefficients from the unitarity of the transmission probability considering the three-terminal model in the linear-response regime. 
This bound prevented the power from being finite with the Carnot efficiency. 
We also note that not Onsager's reciprocal theorem but the unitarity guarantees the non-negativity of the entropy production in the mesoscopic transport theory.
 
Brandner and Seifert \cite{brandner2013multi} argued the attainability of the Carnot efficiency at a finite power in a multiterminal model when the number of probes (a probe is a special reservoir in which we set the temperature and chemical potential so that the net particle and heat currents flowing into the probe may be zero) is infinite, but they themselves denied it later in Ref.~\cite{brandner2015bound} because of a numerically conjectured inequality for the Onsager coefficients.
From our point of view, we presume that the inequality found in Ref.~\cite{brandner2015bound} is based on the unitarity.

To summarize, Benenti \textit{et al}.~\cite{benenti2011thermodynamic} must have missed additional constraints that prevent the Carnot efficiency at a finite power.
In mesoscopic thermoelectric heat engines with the Landauer-B\"uttker formula, the constraint is the unitarity of the transmission probability.
It is an open problem as to what constraints prevent the Carnot efficiency at a finite power in general cases. 

\section{Conclusion}
In this paper, we derived the heat currents coming into or from the central system under reasonable assumptions, with which we can produce Eqs.~\eqref{eq:IV} and \eqref{eq:effc3} as those of cyclic heat engines.  
The same expressions have been used in the previous researches without detailed explanations. These heat currents gave us interesting results when we applied them to the \textit{mesoscopic} thermoelectric heat engine within the Landauer-B\"uttiker formalism.
We found that the heat currents in this engine Eq.~\eqref{eq:jql} correctly give the non-negativity of the entropy production. 
We also found that the \textit{only} transmission probability to achieve the Carnot efficiency is the delta-like function Eq.~\eqref{eq:tau} and that the unitarity of the transmission probability guarantees the Carnot efficiency at zero power.

It will be interesting to incorporate electron-electron interactions to our theory. If the interactions are elastic, Eq.~\eqref{eq:JQ} is still valid because it does not break our assumptions; we may even be able to derive microscopic expressions extending the Landauer-B\"uttiker formalism as in Refs.~\cite{nishino2009exact, nishino2015universal}. Using Christen and B\"uttiker's nonlinear scattering theory, in which electron-electron interactions are treated as mean-field charging effects \cite{buttiker1993capacitance, christen1996gauge, meair2013scattering, lopez2013nonlinear,hwang2013magnetic,whitney2013thermodynamic,whitney2013nonlinear,meair2013scattering, lopez2013nonlinear,sanchez2013scattering}, we can construct the same formalism by replacing $\tau(\epsilon)$ with $\tau(\epsilon, \TL,\TR,\muL,\muR)$.

It is also interesting to look for stronger bounds than the second law of thermodynamics in some systems, which make it impossible to reach the Carnot efficiency. For a system with many probes, for example, especially under the broken time-reversal symmetry, the transmission probability may not have enough degrees of freedom to be in the form of  Eq.~\eqref{eq:tau}.

We finally mention the possibility of experimental realization of the mesoscopic heat engine.
It may be easy to make the setup of the steady-state heat engine, particularly the mesoscopic thermoelectric heat engine, thanks to the improvement of experimental techniques in mesoscopic transport systems. 
The results in this paper can be used from quantum point contacts to one-dimensional nanowires, as well as to cold atoms \cite{ brantut2013thermoelectric} if we utilize the Landauer-B\"uttiker formula. 
A possible difficulty is how we experimentally observe the heat currents and the efficiency of the heat engine. This problem may be solved in the near future because the technique of observing the energy current has been improved recently \cite{jezouin2013quantum}.
 
K.Y.~is supported by Advanced Leading Graduate
Course for Photon Science (ALPS), the University of Tokyo.
N.H.~is supported by Kakenhi Grants No.~15K05200, No.~15K05207 and No.~26400409 from Japan Society for the Promotion of Science.

%

\end{document}